\documentclass{PoS}
\usepackage[T1]{fontenc}
\usepackage{newtxtext,newtxmath}
\usepackage{ae,aecompl}
\usepackage{graphicx}	
\usepackage{amsmath}	
\usepackage{amssymb}	
\usepackage{footmisc}
\usepackage{url}
\usepackage{subfig}
\usepackage{algpseudocode}
\usepackage[noend]{algorithm}
\usepackage{float}
\algtext*{EndIf}

\title{Differential Observation Techniques for the SZE-21cm and radio sources}

\ShortTitle{Differential Observation Techniques for the SZE-21cm and radio sources}

\author{\speaker{Charles Mpho Takalana}\\
	University of the Witwatersrand\\
	E-mail: \email{mtakalana@ska.ac.za}}

\author{Paolo Marchegiani\\
	University of the Witwatersrand\\
	E-mail: \email{Paolo.Marchegiani@wits.ac.za}}

\author{Sergio Colafrancesco\thanks{Deceased - 30 September 2018}\\
	University of the Witwatersrand\\
	E-mail: \email{Sergio.Colafrancesco@wits.ac.za}}

\abstract{The SZE-21cm has been proposed as an alternative probe for the Dark Ages (DA) and the Epoch of Reionization (EoR). The effect is produced when photons of the 21cm background are inverse Compton (up-)scattered by electrons residing in hot plasma of cosmic structures and can be studied through differential observations of the large-scale structure, towards and away a region of the radio sky containing the cosmic structure of interest. This work makes use of the 21cmFAST code to simulate low frequency radio observations of a galaxy cluster from which we then extract the SZE-21cm signal. We further explore uses of the differential technique to radio observations of active radio galaxies using radio observations from the MWA GLEAM survey. The differential techniques this work studies enable us to extract the SZE-21cm signal from simulated galaxy cluster observations, which can then be used to obtain the global features of the 21cm signal during the DA and EoR. The technique shows further benefits in source extraction and characterisation, and feature enhancement in radio observations particularly for low-surface brightness and extended radio sources.}

\FullConference{High Energy Astrophysics in Southern Africa - HEASA2018\\
		1-3 August, 2018\\
		Parys, Free State, South Africa}

\begin{document}

\section{Introduction}
\noindent
Differential observation analysis techniques constitute of acquiring measurements on a target or cosmic structure of interest and the associated background. Then, calculating the difference between the target and the background, which can be used to reveal changes in luminosity of the target less the background. Differential analysis of the 21cm background can be used to probe the history and physics of the DA and EoR using a specific form of the Sunyaev Zeldovich Effect (referred to here as the SZE-21cm) produced in galaxy clusters by inverse Compton scattering of the radio background photons by electrons residing in cluster atmospheres \cite{Colafrancesco,SZ,SZ1,SZ2,SZ3}. The global signatures imprinted on the 21cm background at a given frequency as a function of redshift are measurable today in terms of the brightness temperature frequency spectrum relative to the black-body cosmic microwave background (CMB) \cite{Colafrancesco,Cooray,Furlanetto}. Studies of the global signal will provide unique insights into the physics of cosmic neutral gas and the reionization history of the Universe \cite{Shaver}. There are observational challenges that upcoming experiments are expected to face when studying the 21cm cosmological signal, namely, the 21cm radiation is faint of the order of tens of mK \cite{Pritchard} and foreground emission, which is 4-5 orders of magnitude stronger than the cosmological signal coupled with various experimental systematics and biases. Cooray 2006 \cite{Cooray} and Colafrancesco et al. 2016 \cite{Colafrancesco} proposed the SZE-21cm as a probe for the DA and EoR using differential observation analysis techniques and pointed out various advantages the technique presents as compared to other proposed probes which include studying the 21cm fluctuations \cite{Barkana}, and measuring the contrast between the 21cm signal and the bubbles of ionized plasma \cite{Furlanetto}. They specifically pointed out the weak dependence of the differential technique on exact calibration of the observed intensity and confusion from galactic foregrounds. 
\\
\noindent
Colafrancesco et al. 2016 \cite{Colafrancesco} calculated and made predictions for observations of the SZE-21cm taking into account relativistic corrections, which Cooray 2006 \cite{Cooray} had not considered in his initial calculations of the effect. Using simulated observations, Takalana et al. 2018 \cite{Takalana}  demonstrated how the differential techniques could be applied to low frequency radio observations towards and away from galaxy clusters to extract the SZE-21cm and obtain the global 21cm background signal. Takalana et al. 2018 \cite{Takalana} showed how intrinsic foregrounds and variations can be appropriately removed in order to obtain reliable measurements of the SZE-21cm. The work presented here is a follow-up to Takalana et al. 2018 \cite{Takalana} and does further analysis on the results. This work also looks at and shows additional applications for the differential technique which are now populating the fields of extended and low-surface brightness source finding, extraction and characterization in low-frequency radio astronomical surveys.
\\
This paper is organised as follows. Section \ref{SZE21} tackles the scattering of the 21cm background in clusters of galaxies ultimately leading to the SZE-21cm, the section also presents some results and discussions from this work. Section \ref{new} presents the applications of the differential techniques to general low frequency observations focusing particularly on extended sources. Section \ref{conc} presents our conclusion and basis for future work. For work presented in this text we assume {$\Lambda$}CDM-cosmology with parameters: $h$ = 0.673, $\Omega_{m}$ = 0.315 ,  $\Omega_{b}$ = 0.0491, $\Omega_{\Lambda}$ = 0.685, $\sigma_{8}$ = 0.815 and $n_{s}$ =  0.968 \cite{Planck2016, Takalana}. 

\section{Analysis of differential observations of the cosmological radio background: studying the SZE-21cm}
\label{SZE21}

\subsection{Scattering of the 21cm background in galaxy clusters}

\begin{figure}
	\centering
	\includegraphics[width=5.2in]{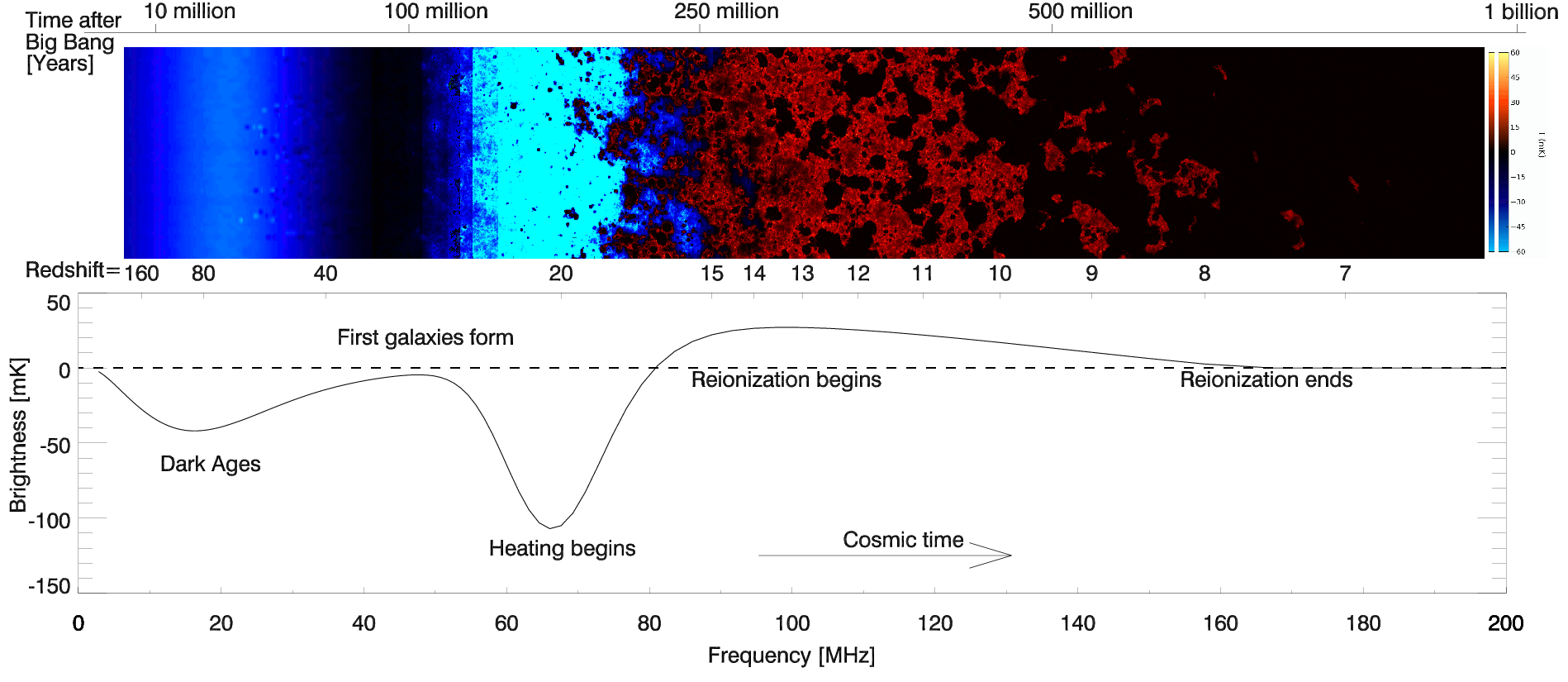}
	\caption{\textit{The 21cm cosmological hydrogen signal. (Top) Time evolution of fluctuations in the 21cm brightness from the DA and to the end of the EoR. (Bottom) Expected evolution of the sky-averaged 21-cm brightness from the DA at redshift 200 to the end of the EoR \cite{Pritchard1}.}} 
	\label{fig:21cm} 
\end{figure}

\noindent
The 21cm line has a rest-frame frequency of 1420 MHz, which is observable as $\nu$ = $1420/(1 + z) MHz$ as a result of the expansion of the Universe, where $z$ is the redshift. The intensity of the 21cm background signal from the DA and EoR is given as a brightness temperature relative to the Cosmic Microwave Background (CMB) \cite{Bowman}:

\begin{align}
	\hspace*{5mm} \delta T_b \approx 23x_{HI}(z) \left[ \left(\frac{1+z}{10} \right) \left(\frac{0.15}{\Omega_{m}} \right) \right] ^{1/2} \left(\frac{\Omega_{b}h}{0.02} \right) \left[1- \frac{T_{R}(z)}{T_{s}(z)}\right] mK
	\label{eq:eq1}, 
\end{align}

\noindent
where $T_{R}$ is the temperature of the background radiation which we assume to be from the CMB, $x_{HI}(z)$ is the fraction of neutral hydrogen at redshift $z$, $\Omega_{m}$ and $\Omega_{b}$ are the matter and baryon densities, respectively, in units of the critical density for a flat universe, $h$ is the Hubble constant in units of 100 km s$^{-1}$ Mpc$^{-1}$ , $T_S$ is the 21cm spin temperature that defines the relative population of the hydrogen hyperfine energy levels, and the factor of 23 mK comes from atomic-line physics and the average gas density. Figure \ref{fig:21cm} summarises the key features of the 21cm signal showing the relevant cosmic time, frequency, and redshift scales. The reader is referred to other specialised texts \cite{Evoli,ewen,Field,Furlanetto,Muller,Pritchard,Pritchard1, van,Wouthuysen} for detailed theory on the 21cm signal.

\noindent 
The 21cm emission of hydrogen gas appears as a faint diffuse background across the sky. At redshifts relevant to studies of the DA and EoR, the primary source of background for measurement of the 21cm cosmological signal is the CMB. This background interacts with the plasmas hosted by large-scale structures such as galaxy clusters and lobes of radio-galaxies, the interaction causes a spectral distortion to the background radiation, this is known as the SZE-21cm. Colafrancesco et al. 2016 \cite{Colafrancesco} gave a full treatment of the derivation of the SZE-21cm, here we just give an overview of the formalism.
\\
\noindent
The CMB photons we observe today traversed the Universe from the last scattering surface and has interacted with matter along their path through the Universe. Energy injecting processes that occur during the DA and EoR modify the CMB spectrum since it is an interaction between the photons of the CMB and electrons of cosmic structures. The incident CMB radiation spectrum is given by

\begin{align}
\hspace*{5mm}I_{0,st}(x) = 2 \frac{(kT_{CMB})^{3}}{(hc)^{2}} \frac{x^{3}}{e^{x}-1},
\label{eq:eq2}
\end{align}

\noindent
we denote the CMB radiation modified by processes during the EoR and DA by $I_{0,mod}$ which can be written as

\begin{align}
\hspace*{5mm}I_{0,mod}(\nu) = I_{0,st}(\nu) + \delta I(\nu)
\label{eq:eq3}
\end{align}

\noindent
where $\delta I(\nu)$ is the perturbation of the CMB radiation due to processes that occur during the DA and EoR. This is related to $\delta T_{b}$ in equation \ref{eq:eq1} by: $\delta T(\nu)$ =  $\delta I(\nu) c^{2} / 2 k_{B} \nu^{2}$. The SZE-21cm is a photon-electron interaction, which alters the frequency of photons due to the motion of the electrons, hence it is written as the change in intensity given by

\begin{align}
\hspace*{5mm}\Delta I_{mod}(\nu) = I_{mod}(\nu) - I_{0,mod}(\nu)
\label{eq:eq4}
\end{align}

\noindent
where $I_{mod}(\nu)$ is the spectral distortion of the CMB due to the SZE, and is given by

\begin{align}
\hspace*{5mm}I(x) = \int_{-\infty}^{+\infty} I_{0}(xe^{-s}) P(s) ds
\label{eq:eq5}
\end{align}

\noindent
$I_{0}$ is incoming radiation spectrum (that can be $I_{0,st}$ for the standard SZE and $I_{0,mod}$ for the SZE-21cm), $x$ is the normalised frequency given by the relation ${x = \frac{h\nu}{kT_{CMB}}}$ and $P(s)$ is the photon redistribution function, which computed by convolving an electron distribution function with the redistribution function for a single electron. $s$ = ln$\frac{\nu^{\prime}}{\nu}$, where  $\nu^{\prime}$  and $\nu$ are the photon frequencies after and before the scattering, respectively. \cite{Birkinshaw1999, Ensslin}. Equation \ref{eq:eq4} can be rewritten in terms of the brightness temperature as

\begin{align}
\hspace*{5mm}\Delta T_{mod}(\nu) = T_{mod}(\nu) - T_{0,mod}(\nu).
\label{eq:eq6}
\end{align}

\subsection{Simulated differential observation of the SZE-21cm}

\noindent
In Takalana et al. 2018 \cite{Takalana} we simulated the cosmological 21cm radio background using 21cmFAST \cite{Mesinger}, which employs semi-numerical approaches to produce large-scale simulations of the 21cm brightness temperature described by equation \ref{eq:eq1}. For this work we use the three dimensional 30 $h^{-1}$ comoving Mpc simulation cubes as in Takalana et al. 2018 \cite{Takalana}, from redshift z = 6.89 up to redshift z = 70, with an initial grid of $460^{3}$ voxels smoothed down to a $115^{3}$ grid. We obtain cubes of the CMB modified by neutral hydrogen ($T_{0,mod}$) during the DA and EoR  as given in equation \ref{eq:eq3}. Takalana et al. 2018 outlines the procedure used to account for bright sources and temperature variations in the cubes. The variations were of order  $10^{-1}$ -  $10^{-3}$ and created a great concern as they would lead to incorrect measurements for the SZE-21cm when conducting our differential observations. The SZE-21cm is a CMB effect and the variations in the cubes were made to be similar to the CMB primodial fluctuations of order $10^{-5}$, this was achieved using a pixel replacement operation where the pixels causing high variations were replaced with the cube mean temperature. With assistance of equation \ref{eq:eq5} the mean value of the $T_{0,mod}$ cube is used to compute the spectral distortion of the CMB due to the SZE ($T_{mod}(\nu)$) at a cluster with temperature of 10 keV, optical depth of {$10^{-3}$} and a radius of 1.5 Mpc (typical of rich clusters where the SZE has been observed). The simulated cluster takes on a shape and size similar to the Coma cluster and the SZE we consider at the cluster is due only to the thermal effect. We insert the cluster ($T_{mod}(\nu)$) in the cube, take a slice of the cube at the cluster and another slice away from the cluster where the slice is only the background ($T_{0,mod}$). Using Equation \ref{eq:eq6} we obtain the SZE-21cm at the cluster ($\Delta T_{mod}(\nu)$), unlike in Takalana et al. 2018 where we assumed an isotropic cluster signal here we model the SZE-21cm signal assuming the cluster gas is well-described by an isothermal $\beta$-model, and is in hydrostatic equilibrium \cite{Cavaliere}. The radial brightness temperature profile of the SZE-21cm takes a simple analytic form:

\begin{align}
\hspace*{5mm}\Delta T_{mod} = \Delta T_{i,mod} \bigg(1 + \frac{\theta^{2}}{\theta^{2}_{c}}\bigg)^{(1-3\beta)/2}   ,
\label{eq:eq7}
\end{align}

\begin{figure}
	\centering
	\includegraphics[width=6.in]{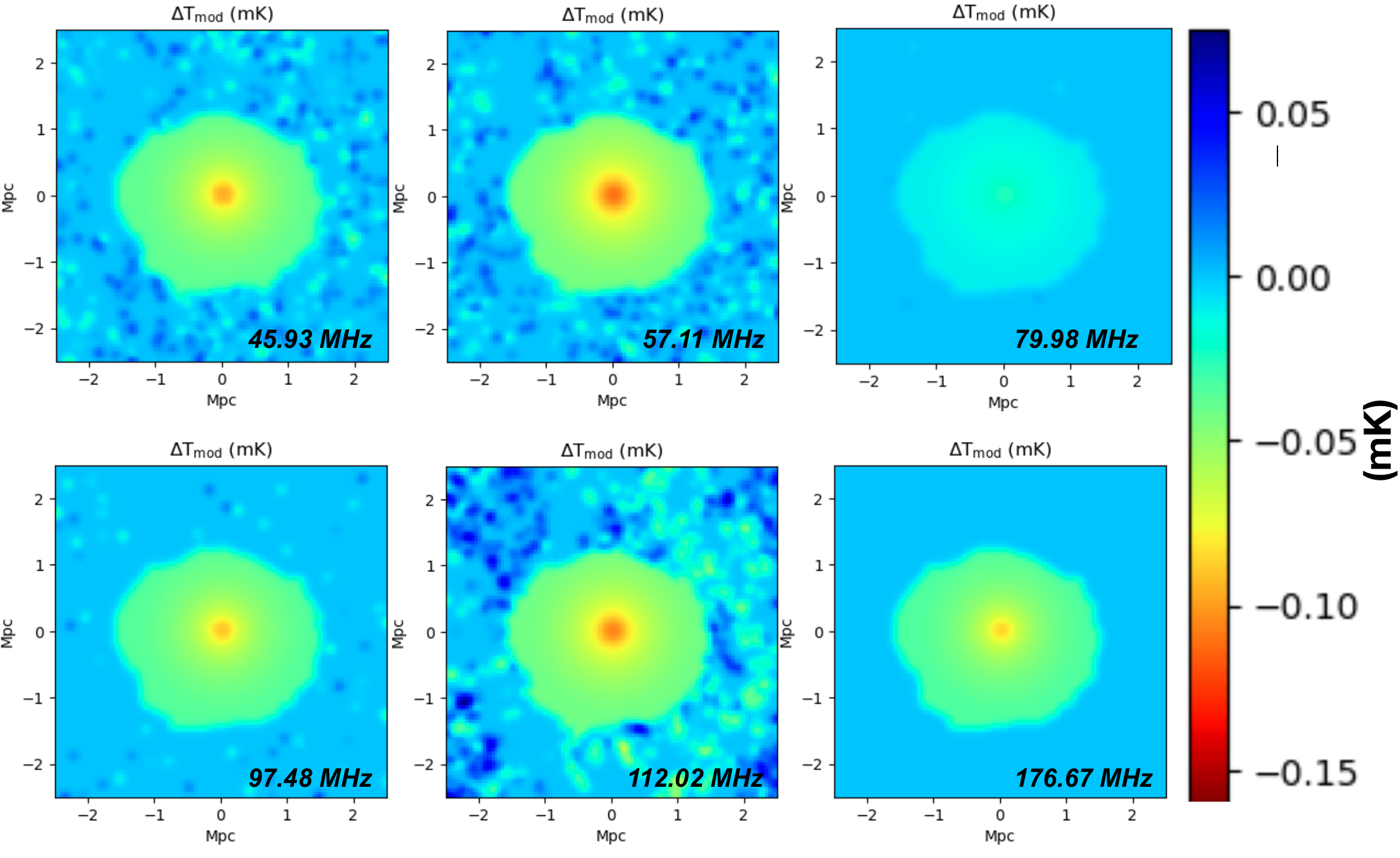}
	\caption{\textit{Image of the SZE-21cm obtained at the cluster using the differential observation procedure for a cluster with a radius of 1.5 Mpc, a temperature of 10 keV, and optical depth of 10$^{-3}$ at: 45.93 MHz, 57.11 MHz, 79.98 MHz, 97.48 MHz, 112.02 MHz and 176.67 MHz.}} 
	\label{fig:sze21cmimi} 
\end{figure}

\noindent
where $\Delta T_{i,mod}$ is the central temperature decrement, and $\theta_{c}$ = $r_{c}$/$D_{A}$ is the core radius divided by the angular-diameter distance. We show sample images of the cluster region in figure \ref{fig:sze21cmimi} at various frequencies, we focus on a 5 Mpc region where the cluster is located. The noise in the simulated observation is due to sources in the background image. The SZE-21cm was measured in each cluster image following the differential procedure (equation \ref{eq:eq6}) at frequency range 20 to 180 MHz and used to construct the spectrum in figure \ref{fig:sze21cm}.

\subsection{Discussion: Differential analysis of the SZE-21cm}

\begin{figure}
	\centering
	\includegraphics[width=4.5in]{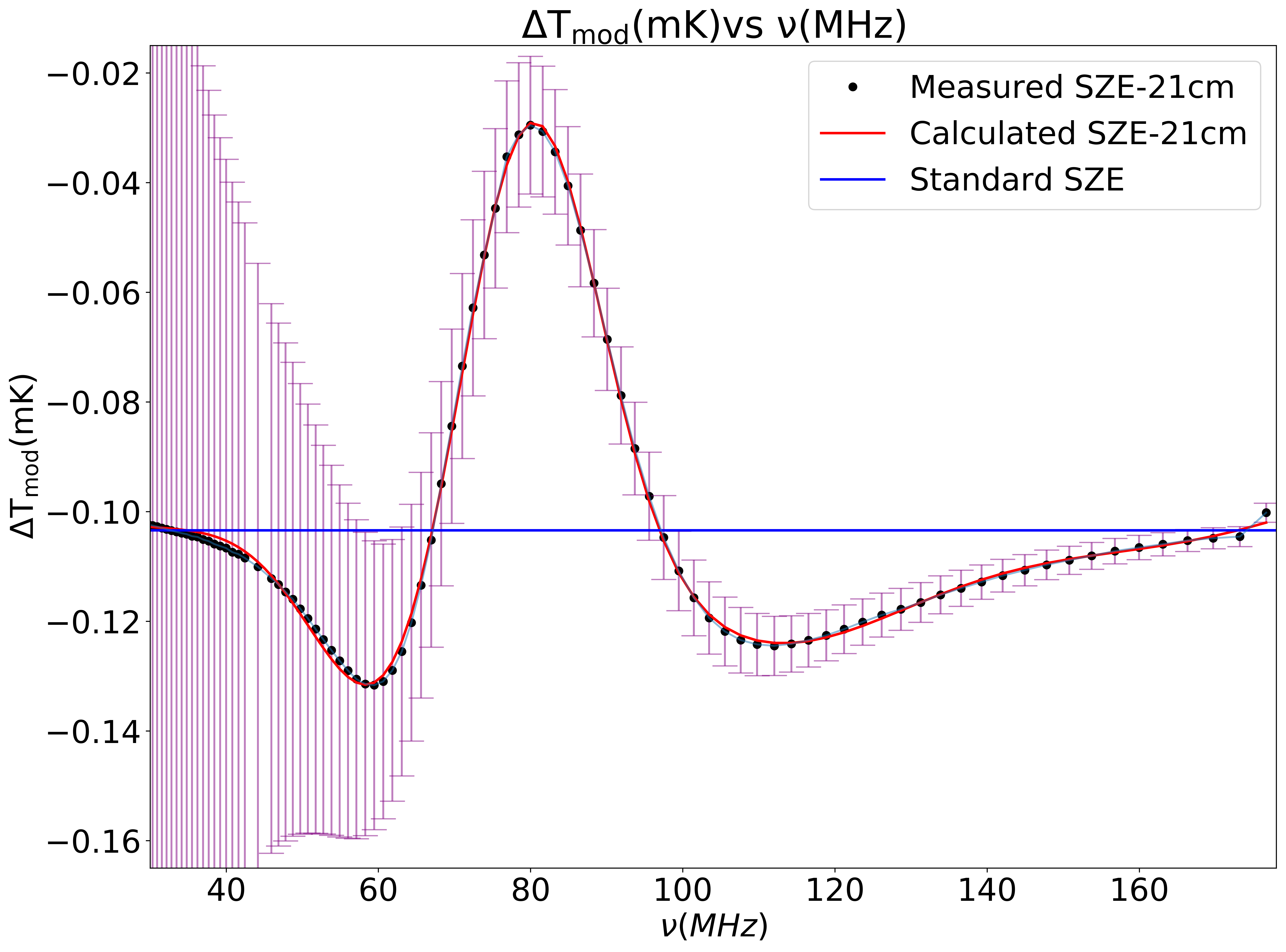}
	\caption{\textit{The SZE-21cm measured following the differential procedure for a cluster of 10 keV, an optical depth of {$10^{-3}$}. The red line is the best-fit line of theoretically calculated data points (Colafrancesco et al. 2016 \cite{Colafrancesco}), the black dots are points measured from the images and the blue line is the standard SZE. The errors bars assume experimental parameters similar to SKA-low with a system noise temperature of 590 K at 130 MHz, and a scaling of $\nu^{-2}$ to other frequencies, a 5 MHz bandwidth synthesized beam of $5^{\prime}$, and a year-long integration. \cite{Cooray}}}
	\label{fig:sze21cm} 
\end{figure}

\noindent
Figure \ref{fig:sze21cmimi} shows that differential observations of low redshift galaxy clusters ($z$ $\sim$ 0.01) can be achieved at low frequencies and hence make it feasible to pursue observations with current and upcoming low frequency radio interferometers, which are not themselves sensitive to the mean 21cm background spectrum as anisotropy observations inherently remove the mean signal on the sky. The SZE-21cm spectrum calculated by this work and shown in figure \ref{fig:sze21cm}, which is the brightness temperature difference towards and away from a cluster, promises a possibility to use the modification to the cosmic 21cm background frequency spectrum by scattering via intervening electrons in galaxy clusters to indirectly establish global features in the mean 21cm background spectrum. This will provide unique insight into the physics of cosmic neutral gas and help us understand the reionization history of the Universe. 

\noindent
The SZE-21cm spectrum in figure \ref{fig:sze21cm} has a decrement (40-60 MHz) followed by an increment (60-80 MHz), and another decrement (80-110 MHz). Looking at figure \ref{fig:sze21cmimi} we see that these decrements and the increment are evident in the cluster SZE-21cm images, the difference can be observed evolving as a function of frequency. Clearly, the amplitudes and the crossover frequencies are sensitive to exact details of the process occurring during the DA and EoR, this is evident as the features in the SZE-21cm spectrum (figure \ref{fig:sze21cmimi}) corresponds to what is expected to be the 21cm background global signal (figure \ref{fig:21cm}). To demonstrate the possibility of the SZE-21cm being observed with upcoming interferometers we plot the expected errors on the brightness temperature assuming the specifications of the SKA-low\footnote{\url{http://www.skatelescope.org}} array \cite{Cooray,Dewdney}. For interferometric observations, the pixel noise in the synthesized beam is given by 

\begin{align}
\hspace*{5mm}\Delta T = 0.03 mK \bigg(\frac{A}{A_{SKA}}\bigg)^{-1} \bigg(\frac{\Delta \nu}{10 MHz}\bigg)^{-1/2}  \bigg(\frac{t_{obs}}{1 yr}\bigg)^{-1/2} \bigg(\frac{T_{sys}}{590 K}\bigg) \bigg(\frac{\nu}{130 MHz}\bigg)^{-2} \bigg(\frac{\Delta \theta}{5^{\prime}}\bigg)^{-2},
\label{eq:eq8}
\end{align}

\noindent
we assume an array with a system temperature ($T_{sys}$) of 590 $K$ at 130 MHz, with $T_{sys} \propto \nu^{-2}$, a bandwidth of 10 MHz for observations, an integration time of 1 year, and a collecting area of A $\sim$ $1km^{2}$. In this case the errors shown in Figure \ref{fig:sze21cm} are those using a single cluster, in Takalana et al (in prep) we show the improvement by averaging over signals towards multiple clusters. 
\\
\noindent
The stacking procedure will require us to have prior knowledge on the cluster optical depth profile and the electron temperature, which will be obtained from existing SZE observations related to the CMB spectrum alone conducted at high radio frequencies \cite{Cooray}. This stacking technique will also be key in addressing and reducing the effect of cosmic variance, with this, we did not include effects of cosmic variance in our calculations. Between tens and hundreds of clusters may be observed in these fields with adequate mass to produce measurable signal of SZE-21cm. Probing of the 21cm background signal with the SZE-21cm at high redshift will not only measure these large scales but also have enough modes to reduce cosmic variance on scales of interest. We expect that SKA1-Low will have fields of view of 7 deg$^{2}$ at 220 MHz (corresponding to 380 comoving Mpc at $z$ = 5.5) to 133 deg$^{2}$ at 50 MHz (corresponding to 2.3 comoving Gpc at $z$ = 27), basically changing as $(1+z)^{2}$ \cite{Pritchard2}. With SKA2 in the pipelines we expect to see further improvements.
\\
\noindent
We have already brought it to the readers attention that differential observational measurements of the SZE-21cm are less affected by contamination from background and foreground emission on scales larger than a galaxy cluster size \cite{Cooray,Colafrancesco}. However, synchrotron radio emission from the host galaxy cluster is a possible source of contamination, Colafrancesco et al. 2016 \cite{Colafrancesco} has shown that this contamination should decrease for objects at large distances as synchrotron emission varies with luminosity distance, whereas the SZE is redshift independent and does not change with redshift and distance of the source. It remains of importance that we study both contributions in order to be able separate them. For this reason Colafrancesco et al 2016 \cite{Colafrancesco} has proposed that a detailed spectral fit of the data be made to separate residual synchrotron and SZE-21cm signals.

\section{Differential analysis techniques for radio observations}
\label{new}

\subsection{Procedure for differential analysis for radio source observations}

\noindent
A procedure similar to that used in the calculation of the SZE-21cm can be adapted to existing continuum radio observations. In this section we show how this can be done and present a preliminary pipeline, which is similar to those that exist for optical experiments (e.g. Levesque \& Lelievre 2007 \cite{Levesque}). The work in this section is rooted in optical pipelines as differential analysis for radio observation continuum surveys is new and a lot can be learnt from optical pipelines with similar goals, there is a scarcity of literature in this regard. Using a stacked background for this procedure helps ensure that the variation we are measuring effectively comes from the main target, and not from the selected background. Equation \ref{eq:eq6} represents how the differential analysis technique can be used to obtain the SZE-21cm which probes the EoR and DA. For the case of the SZE-21cm our main target is a galaxy cluster with spectral distortion due to the SZE ($ T_{mod}(\nu)$) and the background is the incident CMB radiation field ($T_{0,mod}(\nu)$) modified during the DA and EoR. Subtracting the background from the main target we obtain the modification which is the SZE-21cm.

\begin{figure}
	\centering
	\includegraphics[width=5.3in]{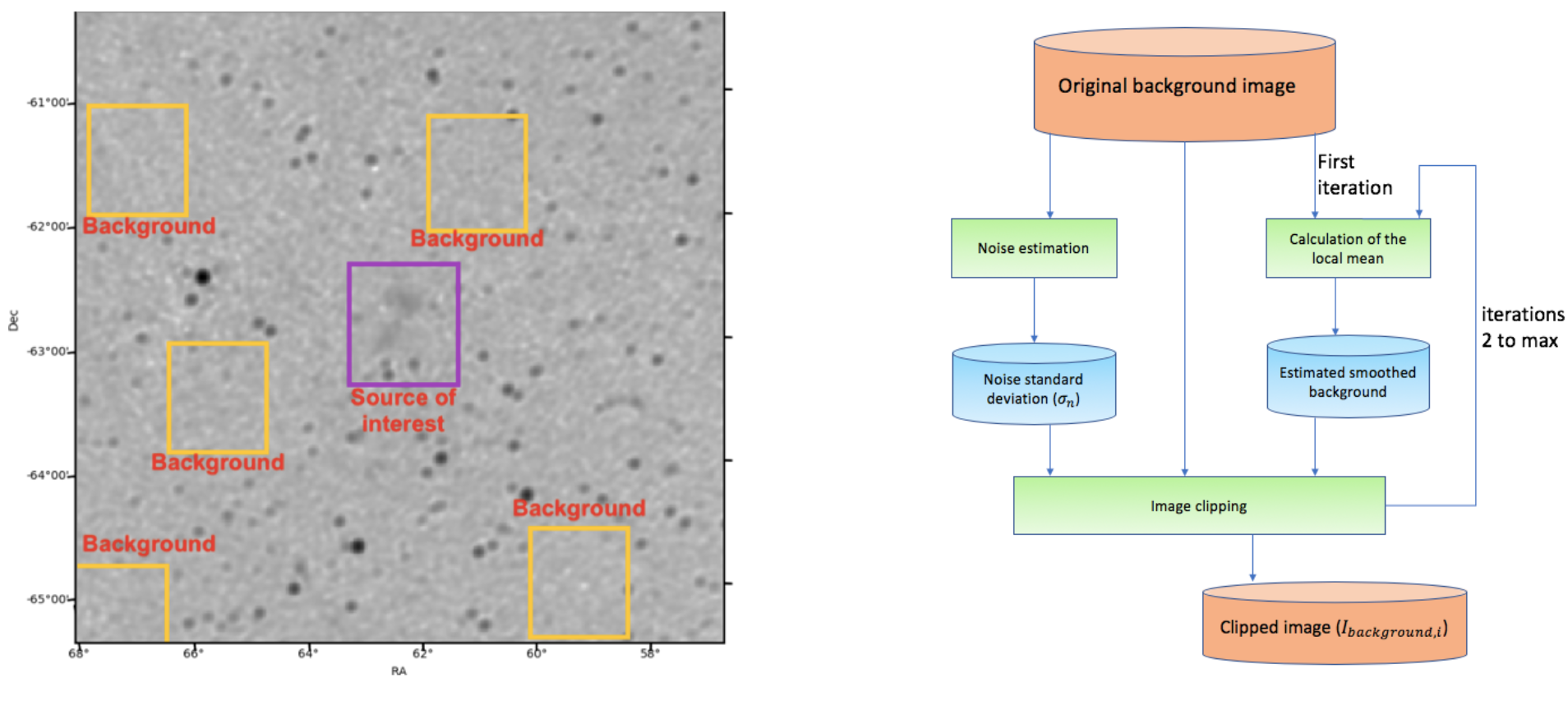}
	\caption{\textit{\textbf{Left:} GLEAM image cut-out with the main target in the purple tile and the five background tiles in yellow. \textbf{Right:} The background calculation pipeline.}}
	\label{fig:pipeline} 
\end{figure}

\noindent
We test our differential technique on images from the GaLactic and Extragalactic MWA (GLEAM) survey carried out with the Murchison Widefield
Array (MWA). The GLEAM survey mapped the southern sky below a declination of +25$\deg$ \cite{Wayth}. These observations consist of roughly an hour of integration per pointing. We selected a patches in the sky from image cut-outs at an observing bandwidth of 147-154 MHz for NGC 1534 and 208-216 MHz for Abell 3565. We identified sources in each image and chose these as our main targets, then selected five background regions with minimal bright sources in each tile, see figure \ref{fig:pipeline} ($left$). We assume that the background tiles are relatively smooth and any signal in the tiles is very sharp, then using an iterative process we attempted to refine the background. The iterations stop when no more improvements can be made to the background tile or when the residual error is less than the noise level. This iterative process is illustrated in Figure \ref{fig:pipeline} ($right$).

\noindent
The signal is almost zero everywhere in the background tiles, and any signal above this background is a first estimation of a source signal. To obtain a better background estimation, the signal must be attenuated before estimating the background again. This is done by clipping the background image at 2$\sigma_{n}$, the pixels affected by signal are replaced by the local mean while the background pixels are preserved. At each iteration any signal is clipped, which causes the resultant image to converge toward the true background. The clipped image where no further improvements can be made to the background is the final background. We follow the same iterative process for the five background tiles and then calculate a mean background image by,

\begin{equation}
\hspace*{5mm}I_{background}  = \frac{1}{K} \sum_{i=1}^{K} I_{{background}_{i}}.
\label{eq:eq9}
\end{equation}

\noindent
The calculated mean background ($I_{background}$) is then subtracted from our main target image with the extended radio source through,

\begin{align}
\hspace*{5mm}\Delta I = I_{source} - I_{background}.
\label{eq:eq10}
\end{align}

\begin{figure}
	\centering
	\includegraphics[width=5.5in]{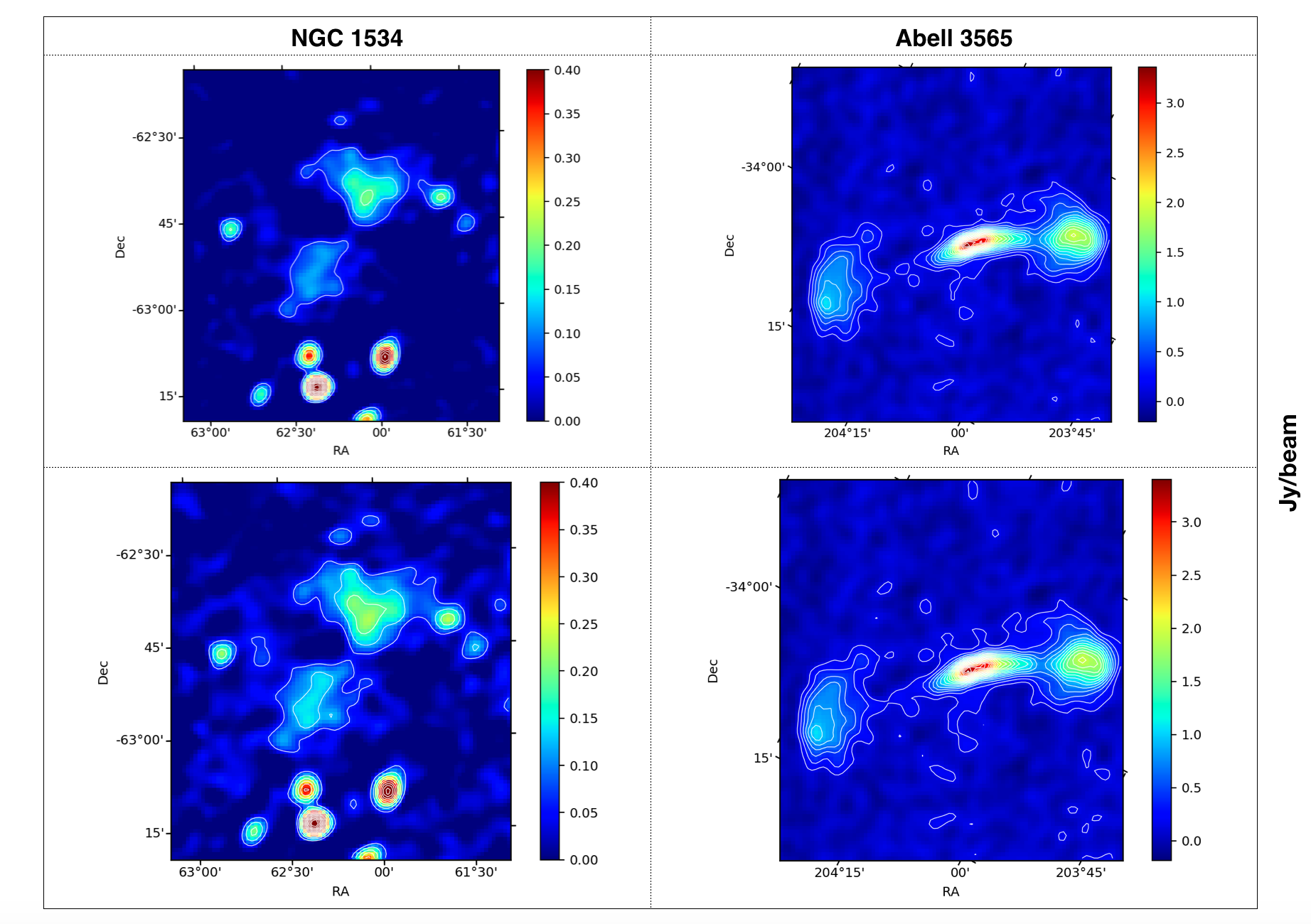}
	\caption{\textit{\textbf{Top:} The main target GLEAM images of NGC 1534 and Abell 3565 before subtracting the background. \textbf{Bottom:} The main target images of NGC 1534 and after subtracting the calculated background.}}
	\label{fig:obser} 
\end{figure}

\noindent
Figure \ref{fig:obser} shows us the main targets (NGC 1534 and Abell 3565) before ($top$) and after ($bottom$) the background subtraction. We do not focus on the full analysis of the main target but only look at the effectiveness of the differential technique. 

\subsection{Discussion: Differential analysis  for low frequency radio emission}

\noindent
Comparing the before and after images in figure \ref{fig:obser} we find that features that were not visible in the original target image are now visible, one can observe these differences by changes in the radio contours. The contours assist us in tracing the extent of radio emission from radio sources in the field of interest. We are able to connect or attribute the emission to a parent source. Where a question arises with regards to radio emission being from a single source or multiple sources, the examples in figure \ref{fig:obser} show a connection of the brighter sources of emission, which lead us to resolve that the radio sources are indeed an extended radio emission. For NGC 1534 the radio emission appears to originate from two non-related sources, however, following the differential operation we observe a connection and enhancement in the extent of the emission. In the case of Abell 3565, the initial image shows a definite connection between the right and central radio emission. The emission on the left seems to be related, however, it is only following the differential operation that we observe a connection of all three sources of the extended radio emission. The signal-to-noise ratio improves by a factor of about 1.3 and 1.5 for NGC 1534 and Abell 3565, respectively, as compared to the original target images. The complete pipeline for the differential techniques for radio emission incorporates the background calculation for the individual tiles, obtaining the mean background, and subtracting the mean background image from the target source image. Development of the pipeline continues but the method has shown some of the benefits it can deliver.

\section{Conclusion}
\label{conc}

\noindent
A measurement of the global 21cm brightness temperature spectrum in figure \ref{fig:21cm} is a goal for current and upcoming experiments. However, these are faced with great observational challenges. This work has shown that observational measurements in the form of modifications to the black body CMB spectrum will provide unique insights into the physics of cosmic neutral gas and the reionization history of the Universe using the SZE-21cm. The Experiment to Detect the Global EoR Signature (EDGES) \cite{Bowman} has already claimed the detection of an absorption feature of the global 21cm background signal centered at 78 MHz. The SZE-21cm will be a great test-bed for the validity of such results as we may expect more in the near future. The differential analysis technique in this work has also shown that it enables enhancement of features in continuum radio images, this may have a wide range of applications in future.

\noindent
\acknowledgments 
\noindent
This research was supported by the South African Radio Astronomy Observatory, which is a facility of the National Research Foundation, an agency of the Department of Science and Technology. C.M. $\&$ P.M. acknowledge the support, guidance and supervision of Prof. Sergio Colafrancesco in this research, may he rest in peace.

\end{document}